\documentclass{article}
\usepackage{cite}
\usepackage{graphicx}
\usepackage{dcolumn}

\input{tcilatex}
\begin{document}

\date{}
\title{Comment on: ``On the Dirac oscillator subject to a Coulomb-type central
potential induced by the Lorentz symmetry violation''. }
\author{Francisco M. Fern\'{a}ndez\thanks{%
fernande@quimica.unlp.edu.ar} \\
INIFTA, DQT, Sucursal 4, C.C 16, \\
1900 La Plata, Argentina}
\maketitle

\begin{abstract}
We analyze recent results on a Dirac oscillator. We show that the truncation
of the Frobenius series does not yield all the eigenvalues and
eigenfunctions of the radial equation. For this reason the eigenvalues
reported by the authors are useless and the prediction of allowed oscillator
frequencies meaningless.
\end{abstract}

In a recent paper Vit\'{o}ria and Belich\cite{F20b} investigated the
relativistic oscillator model for spin-1/2 fermionic fields, known as the
Dirac oscillator, in a background of breaking the Lorentz symmetry governed
by a constant vector field inserted in the Dirac equation by non-minimal
coupling. The authors proposed two possible scenarios of Lorentz symmetry
violation which induce a Coulomb type potential. They obtained the
relativistic energy profile for the Dirac oscillator finding that the
frequency of the Dirac oscillator is determined by the quantum numbers of
the system and the parameters that characterize the scenarios of Lorentz
symmetry violation.

By a suitable separation of variables the authors arrived at an eigenvalue
equation for the radial part of the solution and applied the Frobenius
method. Since the coefficients of the expansion satisfy a three-term
recurrence relation the authors forced a truncation of the series in order
to obtain exact eigenfunctions and eigenvalues. In this Comment we analyze
the effect of the truncation method just mentioned on the physical
conclusions drawn by the authors.

In the first case the authors derived the following eigenvalue equation
\begin{eqnarray}
&&\frac{d^{2}\psi _{s}}{d\varrho ^{2}}+\frac{1}{\varrho }\frac{d\psi _{s}}{%
d\varrho }-\frac{\iota _{s}^{2}}{\varrho ^{2}}\psi _{s}-\frac{\beta }{%
\varrho }\psi _{s}-\varrho ^{2}\psi _{s}+W\psi _{s}=0  \nonumber \\
&&W=\frac{\alpha ^{2}}{m\omega },\;\iota _{s}^{2}=\left[ l+\frac{1}{2}%
(1-s)\right] ^{2}-(ag\lambda )^{2},\;\beta =\frac{2ag\lambda \mathcal{E}}{%
\sqrt{m\omega }}  \nonumber \\
&&\alpha ^{2}=\mathcal{E}^{2}-m^{2}+2m\omega \left( l+\frac{1}{2}\right)
s+m\omega  \label{eq:eig_eq_CH}
\end{eqnarray}
where $l=0,\pm 1,\pm 2,\ldots ,$ $a$ is a constant, $\lambda $ is a
parameter associated to the linear electric charge distribution, $\omega $
is the oscillator frequency, $m$ a mass and $\mathcal{E}$ the energy. The
authors simply stated that $\hbar =c=1$, although there are efficient and
rigorous ways of obtaining suitable dimensionless equations\cite{F20}. From
a truncation method that we discuss below the authors obtained the energies $%
\mathcal{E}_{1,l,s}$ in terms of an angular frequency $\omega _{1,l,s}$ that
depends on the quantum numbers.

In the second case the authors arrived at the following eigenvalue equation
\begin{eqnarray}
&&\frac{d^{2}\psi _{s}}{d\varrho ^{2}}+\frac{1}{\varrho }\frac{d\psi _{s}}{%
d\varrho }-\frac{\nu _{s}^{2}}{\varrho ^{2}}\psi _{s}+\frac{\tau }{\varrho }%
\psi _{s}-\eta \varrho \psi _{s}-\varrho ^{2}\psi _{s}+W\psi _{s}=0,
\nonumber \\
&&\nu _{s}^{2}=\left[ l+\frac{1}{2}(1-s)\right] ^{2},\;\epsilon ^{2}=%
\mathcal{E}^{2}-m^{2}+2m\omega \left( l+\frac{1}{2}\right) s+m\omega
-a^{2}B_{0}^{2}g^{2},  \nonumber \\
&&\tau =\frac{2aB_{0}g(l+1/2)}{\sqrt{m\omega }},\;\eta =\frac{2aB_{0}gs}{%
\sqrt{m\omega }},\;W=\frac{\epsilon ^{2}}{m\omega }  \label{eq:eig_eq_CLH}
\end{eqnarray}
where $B_{0}$ is a constant. Again the authors obtained an energy $\mathcal{E%
}_{1,l,s}$ in terms of an angular frequency $\omega _{1,l,s}$ that depends
on the quantum numbers.

In what follows we will show that the authors' interpretation of the results
obtained from the truncation method are meaningless from a physical point of
view. In order to facilitate the discussion we just focus on the eigenvalue
equation
\begin{equation}
\psi ^{\prime \prime }(x)+\frac{1}{x}\psi (x)-\frac{\gamma ^{2}}{x^{2}}\psi
(x)-\frac{a}{x}\psi (x)-bx\psi (x)-x^{2}\psi (x)+W\psi (x)=0,
\label{eq:eig_eq}
\end{equation}
where $\gamma $, $a$ and $b$ are real model parameters that have nothing to
do with the parameter in equations (\ref{eq:eig_eq_CH}) and (\ref
{eq:eig_eq_CLH})). This eigenvalue equation, which is a generalization of (%
\ref{eq:eig_eq_CH}) and (\ref{eq:eig_eq_CLH}), has square-integrable
solutions
\begin{equation}
\int_{0}^{\infty }\left| \psi (x)\right| ^{2}x\,dx<\infty ,
\label{eq:bound_states}
\end{equation}
for particular values (allowed values) of the eigenvalue $W$. Since the
behaviour at origin is determined by $\gamma ^{2}/x^{2}$ and the behaviour
at infinity by the harmonic term $x^{2}$ then we conclude that there are
bound states for all $-\infty <a,b<\infty $. Therefore, the eigenvalues $%
W(a,b)$ are surfaces in the three-dimensional $a\,b\,W$ space. They are
continuous functions of $a$ and $b$ that satisfy the Hellmann-Feynman theorem%
\cite{F39}
\begin{equation}
\frac{\partial W}{\partial a}=\left\langle \frac{1}{x}\right\rangle >0,\;%
\frac{\partial W}{\partial b}=\left\langle x\right\rangle >0.  \label{eq:HFT}
\end{equation}

In what follows we apply the Frobenius method to the eigenvalue equation (%
\ref{eq:eig_eq}) by means of the ansatz
\begin{equation}
\psi (x)=x^{s}\exp \left( -\frac{b}{2}x-\frac{x^{2}}{2}\right)
H(x),\;H(x)=\sum_{j=0}^{\infty }c_{j}x^{j},\;s=\left| \gamma \right| .
\label{eq:ansatz}
\end{equation}
The expansion coefficients $c_{j}$ satisfy the three-term recurrence
relation
\begin{eqnarray}
c_{j+2} &=&A_{j}c_{j+1}+B_{j}c_{j},\;j=-1,0,1,2,\ldots ,\;c_{-1}=0,\;c_{0}=1,
\nonumber \\
A_{j} &=&\frac{2a+b\left( 2j+2s+3\right) }{2\left( j+2\right) \left[
j+2\left( s+1\right) \right] },\;B_{j}=\frac{4\left( 2j+2s-W+2\right) -b^{2}%
}{4\left( j+2\right) \left[ j+2\left( s+1\right) \right] }.  \label{eq:TTRR}
\end{eqnarray}
The authors showed that $H(x)$ is solution to a biconfluent Heun equation
but they did not use the properties of this equation and resorted to a
truncation of the Frobenius series. If the truncation condition $%
c_{n+1}=c_{n+2}=0$, $c_{n}\neq 0$, $n=0,1,\ldots $, has physically
acceptable solutions for $a$, $b$ and $W$ then we obtain exact
eigenfunctions because $c_{j}=0$ for all $j>n$. This truncation condition is
equivalent to $B_{n}=0$, $c_{n+1}=0$ or
\begin{equation}
W_{s}^{(n)}=2\left( n+s+1\right) -\frac{b^{2}}{4},\;c_{n+1}(a,b)=0,
\label{eq:trunc_cond}
\end{equation}
where the second condition determines a relationship between the parameters $%
a$ and $b$. On setting $W=W_{s}^{(n)}$ the coefficient $B_{j}$ takes a
simpler form:
\begin{equation}
B_{j}=\frac{2\left( j-n\right) }{\left( j+2\right) \left[ j+2\left(
s+1\right) \right] }.
\end{equation}

\begin{figure}[tbp]
\begin{center}
\includegraphics[width=9cm]{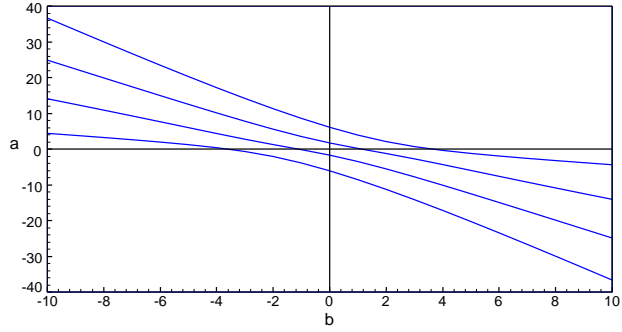}
\end{center}
\caption{Curves $a^{(3,i)}_0$, $i=1,2,3,4$}
\label{Fig:a(b)}
\end{figure}

The second condition $c_{n+1}(a,b)=0$ is a polynomial equation for $a$ and $%
b $ of degree $n+1$ in every variable. From this equation one obtains either
$a_{s}^{(n,i)}(b)$ or $b_{s}^{(n,i)}(a)$, $i=1,2,\ldots ,n+1$, and it can be
proved that all the roots are real\cite{CDW00,AF20}. For example, for $%
n=0,1,2,3$ we have
\begin{equation}
2a+b\left( 2s+1\right) =0,  \label{eq:c(a,b)^0}
\end{equation}
\begin{equation}
4a^{2}+8ab\left( s+1\right) +b^{2}\left( 2s+1\right) \left( 2s+3\right)
-8\left( 2s+1\right) =0,  \label{eq:c(a,b)^1}
\end{equation}
\begin{eqnarray}
&&8a^{3}+12a^{2}b\left( 2s+3\right) +2ab^{2}\left( 12s^{2}+36s+23\right)
-32a\left( 4s+3\right)  \nonumber \\
&&+b^{3}\left( 2s+1\right) \left( 2s+3\right) \left( 2s+5\right) -16b\left(
2s+1\right) \left( 4s+7\right) =0,  \label{eq:c(a,b)^2}
\end{eqnarray}
and
\begin{eqnarray}
&&16a^{4}+64a^{3}b\left( s+2\right) +8a^{2}b^{2}\left( 12s^{2}+48s+43\right)
-640a^{2}\left( s+1\right)  \nonumber \\
&&+16ab^{3}\left( 4s^{3}+24s^{2}+43s+22\right) -128ab\left(
10s^{2}+30s+17\right)  \nonumber \\
&&+b^{4}\left( 2s+1\right) \left( 2s+3\right) \left( 2s+5\right) \left(
2s+7\right) -32b^{2}\left( 2s+1\right) \left( 10s^{2}+45s+47\right)
\nonumber \\
&&+576\left( 2s+1\right) \left( 2s+3\right) =0,  \label{eq:c(a,b)^3}
\end{eqnarray}
respectively. Figure~\ref{Fig:a(b)} shows part of the curves $a_{0}^{(3,i)}$%
, $i=1,2,3,4$.

It follows from the analysis above that the polynomial solutions can be
written as
\begin{eqnarray}
\psi _{s}^{(n,i)}(x) &=&x^{s}\exp \left( -\frac{b}{2}x-\frac{x^{2}}{2}%
\right)
H_{s}^{(n,i)}(x),\;H_{s}^{(n,i)}(x)=\sum_{j=0}^{n}c_{j,s}^{(n,i)}x^{j},\;s=%
\left| \gamma \right| ,  \nonumber \\
n &=&0,1,\ldots ,\;i=1,2,\ldots ,n+1  \label{eq:poly_sol}
\end{eqnarray}
Vit\'{o}ria and Belich\cite{VB20b} only showed eigenvalues for $n=1$ and
overlooked the multiplicity of roots for each value of $n$.

It is worth noticing that the truncation condition does not provide all the
solutions that satisfy equation (\ref{eq:bound_states}) but only polynomial
functions of the form (\ref{eq:poly_sol}) for which the parameters $a$ and $%
b $ exhibit certain polynomial relations like those in equations (\ref
{eq:c(a,b)^0}-\ref{eq:c(a,b)^3}). The reason is that this problem is not
exactly solvable, as Vit\'{o}ria and Belich appear to believe, but
quasi-exactly solvable or conditionally solvable (see \cite
{CDW00,AF20,F20b,F20c} and, in particular, the remarkable review \cite{T16}
and references therein for more details).

It is revealing to compare the eigenvalues given by the truncation method
with the actual eigenvalues $W_{j,s}(a,b)$, $j=0,1,\ldots $ of equation (\ref
{eq:eig_eq}). Since this eigenvalue equation is not exactly solvable\cite
{AF20,T16} we should apply an approximate method. Here, we resort to the
well known Ritz variational method that is known to yield upper bounds to
all the eigenvalues\cite{P68} and, for simplicity, choose the non-orthogonal
basis set of Gaussian functions $\left\{ \varphi _{j,s}(x)=x^{s+j}\exp
\left( -\frac{x^{2}}{2}\right) ,\;j=0,1,\ldots \right\} $.

\begin{figure}[tbp]
\begin{center}
\includegraphics[width=9cm]{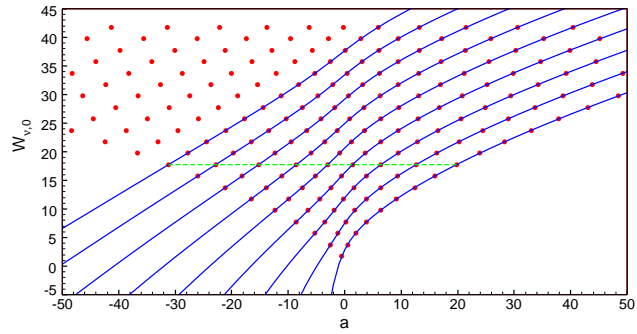}
\end{center}
\caption{Eigenvalues $W_0^{(n)}(a,1)$ from the truncation condition (red
points) and $W_{j,0}(a)$ obtained by means of the variational method (blue
lines)}
\label{Fig:Wb1g0}
\end{figure}

Figure~\ref{Fig:Wb1g0} shows some eigenvalues $W_{0}^{(n)}(b=1)$ given by
the truncation condition (red points) and the lowest variational eigenvalues
$W_{j,0}(a,1)$ (blue lines). We clearly appreciate that the truncation
condition (\ref{eq:trunc_cond}) yields only some particular points of the
curves $W_{j,0}(a,1)$. Therefore, any conclusion drawn from $W_{s}^{(n)}$ is
meaningless unless one is able to organize these eigenvalues properly\cite
{AF20,F20b,F20c}. Vit\'{o}ria and Belich\cite{VB20b} completely overlooked
this fact. The reason is that these authors appear to believe that the only
acceptable solutions to the eigenvalue equation are those with polynomial
factors as in equation (\ref{eq:poly_sol}). The fact is that this kind of
solutions already satisfy equation (\ref{eq:bound_states}) but they are not
the only ones. Notice that the variational method also yields the polynomial
solutions as shown by the fact that the blue lines connect the red points in
Figure~\ref{Fig:Wb1g0}. In order to make the meaning of the eigenvalues $%
W_{s}^{(n)}$ and the associated multiplicity of roots $i=1,2,\ldots ,n+1$
clearer, Figure~\ref{Fig:Wb1g0} shows an horizontal line (green, dashed) at $%
W=W_{0}^{(8)}$ that intersects the curves $W_{j,0}(a,1)$ exactly at the red
points. The most important conclusion of present analysis is that the
occurrence of allowed oscillator frequencies are fabricated by Vit\'{o}ria
and Belich by picking out some isolated eigenvalues $W_{s}^{(n)}$ for some
particular curve $a_{s}^{(n,i)}(b)$. Since there are eigenvalues $%
W_{j,s}(a,b)$ for all real values of $a$ and $b$ then there are bound states
for every positive value of $\omega $ in their equations (\ref{eq:eig_eq_CH}%
) and (\ref{eq:eig_eq_CLH}).

Summarizing: the truncation of the Frobenius series does not yield all the
bound states of the radial eigenvalue equation but only some particular
states of the form (\ref{eq:poly_sol}). These exact solutions only occur for
some relationships between the model parameters $a$ and $b$. For this reason
the eigenvalues reported by Vit\'{o}ria and Belich\cite{VB20b} are
meaningless unless one is able to arrange them carefully as shown in present
Figure~\ref{Fig:Wb1g0} and earlier papers\cite{AF20,F20b,F20c}. The allowed
oscillator frequencies conjectured by those authors are a mere artifact of
the truncation method because there are solutions to equations (\ref
{eq:eig_eq_CH}) and (\ref{eq:eig_eq_CLH}) for all positive values of $\omega
$.

The criticisms outlined in this Comment also apply to other almost identical
papers appeared in the same journal\cite{BB12,BB14,OBB20}.

\begin{thebibliography}{99}
\bibitem{VB20b}  L. L. Vit\'{o}ria and H. Belich, Eur. Phys. J. Plus \textbf{%
135}, 247 (2020).

\bibitem{F20}  F. M. Fern\'{a}ndez, Dimensionless equations in
non-relativistic quantum mechanics, arXiv:2005.05377 [quant-ph].

\bibitem{F39}  R. P. Feynman, Phys. Rev. \textbf{56}, 340 (1939).

\bibitem{CDW00}  M. S. Child, S-H. Dong, and X-G. Wang, J. Phys. A \textbf{33%
}, 5653 (2000).

\bibitem{AF20}  P. Amore and F. M. Fern\'{a}ndez, Phys. Scr. \textbf{95},
105201 (2020). arXiv:2007.03448 [quant-ph]

\bibitem{F20b}  F. M. Fern\'{a}ndez, The rotating harmonic oscillator
revisited, arXiv:2007.11695 [quant-ph].

\bibitem{F20c}  F. M. Fern\'{a}ndez, The truncated Coulomb potential
revisited, arXiv:2008.01773 [quant-ph].

\bibitem{T16}  A. V. Turbiner, Phys. Rep. \textbf{642}, 1 (2016).
arXiv:1603.02992v2

\bibitem{P68}  F. L. Pilar, Elementary Quantum Chemistry (McGraw-Hill, New
York, 1968).

\bibitem{BB12}  K. Bakke and H. Belich, Eur. Phys. J. Plus \textbf{127}, 102
(2012).

\bibitem{BB14}  K. Bakke and H. Belich, Eur. Phys. J. Plus \textbf{129}, 147
(2014).

\bibitem{OBB20}  A. S. Oliveira, K. Bakke, and H. Belich, Eur. Phys. J. Plus
\textbf{135}, 623 (2020).
\end{thebibliography}
\end{document}